\documentclass[11pt]{article}
\usepackage{hyperref}
\pdfoutput=1
\begin{document}
\title{Bursting water balloons}
\author{Hugh M. Lund, Stuart B. Dalziel \\ \vspace{6pt}DAMTP \\  University of Cambridge, \\ Wilberforce Road,
Cambridge, CB3 0WA,\\ United Kingdom}
\maketitle
%% The abstract (in this file, and that submitted as text to arXiv) should
%%include the exact phrase
%% "fluid dynamics video" or "fluid dynamics videos"
\begin{abstract}
In our fluid dynamics video, the rupture of water-filled balloons are shown. The first two experiments depict the impact of 300 mm radius balloons after being dropped from a height of 1 metre onto a flat, rigid surface. The later two videos show 130 mm radius balloons being held underwater, forcibly oscillated at 170 hz and 120 hz respectively, then burst with a pin. The water inside the balloons is coloured with milk powder to aid visualisation. All videos were recorded at 5400 fps and played back at 30 fps.

The initial stages of the impact of a water-filled balloon upon a rigid surface are analogous to those for a water droplet. The impact of small water droplets of constant surface tension upon a rigid boundary has been extensively studied. After impact, the droplet spreads radially and thins, whilst small capillary waves are formed upon the water's free surface. In this way, the droplet's kinetic energy is transferred primarily into surface energy. Once all the kinetic energy is lost, the droplet may then retract due to the pressure gradient created as a result of its deformation. Should the rigid surface be suitably hydrophobic and the impact velocity sufficiently small, it may lift completely off the surface once more, and is said to have \emph{bounced}. 

For a water-filled balloon, the process is very similar. Upon impact, upward-travelling capillary-like waves are created on the membrane, the velocities of which are in approximate agreement with linear potential theory for constant surface tension. Should the balloon membrane remain intact during the impact, it will evolve in a similar way to the surface of a \emph{bouncing} droplet, with the membrane's elastic energy acting analogously to the surface energy of a water droplet. The balloon's initial kinetic energy is then transferred into elastic energy in the membrane, creating pressure forces that cause the balloon to retract. 

However, if the membrane ruptures during impact, any similarities are quickly lost. The increase in elastic energy is no longer transferred back by the pressure field into kinetic energy in the water, but instead is lost to the kinetic energy of the ruptured membrane and to overcome frictional resistance between the latex and the three other phases. The rapid retraction of the membrane then creates a small-scale shear instability on the air/water interface. On a larger scale, as the restoring force for the capillary-like waves is lost while their kinetic energy within the water remains, growth of the interfacial amplitude occurs that may be regarded as a manifestation of the Richtmyer-Meshkov instability. Eventually gravity comes to dominate the flow, leading to the slumping and spreading of the water.

A water-filled balloon that is held, forcibly oscillated then ruptured with a sharp object displays the same three distinct phenomena. In air, the larger scale growth of the interfacial amplitude becomes asymmetric, leading to the formation of so-called bubbles of the less dense, air phase and spikes of the denser, water phase. This is the form expected of Richtmyer-Meshkov instability for an interface between two phases with large density ratio. 

For such a balloon held underwater, the interfacial amplitude grows in a symmetric way. This is a clear distinction with classical Richtmyer-Meshkov instability, as that does not occur on a interface between two phases of the same density. At late time, the displacement of the maximum amplitude of the interface obeys a power law of the form $t^{\theta}$, where $t$ is time and $\theta$ is around $2/3$.\end{abstract}

\end{document}